\documentclass[12pt]{article}

\usepackage{lmodern}
\usepackage{geometry}
\usepackage{fullpage}
\usepackage{microtype}
\usepackage[utf8]{inputenc}
\usepackage[T1]{fontenc}
\usepackage{setspace}
\usepackage[round]{natbib}

\usepackage{hyperref}
\usepackage[title]{appendix}

\usepackage{graphicx}
\usepackage{caption, subcaption}

\usepackage{amsmath, amssymb, amsthm}
\usepackage{algorithm}
\usepackage{algpseudocode}
\usepackage{threeparttable}

\usepackage[autostyle]{csquotes}
\MakeOuterQuote{"}

\theoremstyle{plain}
\newtheorem{thm}{Theorem}[section]
\newtheorem{prop}[thm]{Proposition}
\newtheorem{lemma}[thm]{Lemma}

\newcommand{\E}{\mathbb{E}} 
\newcommand{\vecop}{\operatorname{vec}} \newcommand{\tr}{\operatorname{tr}}
\newcommand{\diag}{\operatorname{diag}} \newcommand{\tsp}{\mathsf{T}}
\newcommand{\Beta}{\mathcal{B}} \newcommand{\R}[1]{\mathbb{R}^{#1}}

\newcommand{\cov}{\operatorname{cov}} 
 \newcommand{\argmax}{\operatorname{argmax}}
\newcommand{\rN}{\mathcal{N}}

\begin{document}

\title{Likelihood-Based Inference with Separable Correlation Matrices}

\author{Karl Oskar Ekvall\\
Department of Statistics, University of Florida\\
\texttt{k.ekvall@ufl.edu}}

\date{}

\maketitle

\begin{abstract}
This paper proposes methods for likelihood-based inference in multivariate linear regressions when the correlation matrix of the responses is separable; that is, it has a Kronecker product structure, but the variances are unrestricted. The methods are enabled by a block-coordinate ascent-like algorithm with closed-form updates that strictly increases the likelihood at every iteration until convergence. In the numerical experiments, the proposed algorithm is 300--2500 times faster than a general-purpose solver, making parametric bootstrap tests of correlation and covariance separability practical. Parameters are identifiable, and standard errors can therefore be obtained from the expected Fisher information, which can be computed efficiently using the Kronecker product structure. Simulations show that the proposed estimator has lower error than both separable covariance and unrestricted estimators when the model holds, and that bootstrap tests maintain nominal size where asymptotic tests fail. An application to dissolved oxygen data from the Mississippi River demonstrates that separable correlation captures location-specific variance patterns that separable covariance cannot.
\end{abstract}

\noindent\textbf{Keywords:} coordinate ascent, Kronecker product, matrix normal distribution, separability, spatio-temporal data, structured covariance estimation

\section{Introduction}
Separable dependence structures arise in many applications, including
multivariate repeated-measures data, longitudinal imaging, diffusion-tensor MRI,
spatio-temporal environmental data, mortality tables, genomics, and
international trade data
\citep{naik2001analysis,george2015selecting,basser2003normal,zapata2022partial,dutilleul1996doubly,fosdick2014separable,allen2012inference,volfovsky2015testing}.
Often, the data are matrix-valued, and then separability says dependence
patterns among rows are similar across columns, and vice versa.

To be more precise, suppose there are $n$ independent $r \times c$ response
matrices $Y_1, \dots, Y_n$; an extension to tensor-valued data is
discussed later. Each $Y_i$ has elements $Y_{ijk}$, $j \in \{1, \dots, r\}$
and $k \in \{1, \dots, c\}$. The focus is on inference on the covariance
structure for observations in the same matrix $Y_i$, which is characterized by the covariance matrix of the vector
of all responses in $Y_i$. Let $\vecop(Y_i) = [Y_{i11}, Y_{i12},
\dots, Y_{irc}]^\top \in \R{rc}$ denote this vector, and $\Sigma =
\cov\{\vecop(Y_i)\}$. The matrix $\Sigma$ is separable if $\Sigma = \Sigma_2 \otimes
\Sigma_1$ for some positive definite $\Sigma_1 \in \R{r \times r}$ and $\Sigma_2
\in \R{c \times c}$. Equivalently, $\Sigma$ is separable if every column of
$Y_i$ has covariance matrix proportional to $\Sigma_1$, the constant of
proportionality being the corresponding diagonal element of $\Sigma_2$, and
every row of $Y_i$ has covariance matrix proportional to $\Sigma_2$.

In some applications, covariance separability is too restrictive, yet estimating
an unstructured covariance matrix is also inappropriate, either because it
ignores the structure of the data or because the number of parameters is
large relative to the sample size. It may then be useful to assume that the
correlation matrix is separable but that the variances need not be. Formally,
\begin{equation} \label{eq:sep_cor}
  \Sigma = D (C_2 \otimes C_1) D,
\end{equation}
where $D = \diag(d_1, \dots, d_{rc}) = \diag(\sqrt{\Sigma_{11}}, \dots,
\sqrt{\Sigma_{rc}})$ is a diagonal matrix of standard deviations, and $C_1$ and
$C_2$ are correlation matrices of dimensions $r \times r$ and $c \times c$,
respectively. The difference from separable covariance is that $D$
need not be separable. Consequently,
the number of parameters is larger than with separable covariance, $rc + r(r -
1)/2 + c(c - 1)/2$ compared to $r(r + 1)/2 + c(c + 1)/2 - 1$, the last $-1$
coming from an identifiability constraint such as $[\Sigma_{2}]_{11} = 1$. Thus,
as $r$ and $c$ grow, both parameter counts are $O(r^2 + c^2)$ and are much
smaller than the $O(r^2c^2)$ parameters needed for an unstructured covariance
matrix.

While separability of correlations has been considered in the spatio-temporal
statistics literature \citep[see][for example]{gneiting2006geostatistical}, the
likelihood-based methods studied here have not. A main contribution is a
block coordinate ascent algorithm with rescaling for maximizing the likelihood.
Thanks to closed-form updates which are fast to compute, the algorithm is very
efficient, often several orders of magnitude faster than off-the-shelf solvers.
It is also stable, increasing the likelihood at every iteration until it reaches
a stationary point.  A full theoretical analysis of the optimization problem,
similar to those available for the separable covariance model
\citep{wiesel2012geodesic,ros2016existence,soloveychik2016gaussian,drton2021existence,derksen2021maximum},
is well beyond the scope of the paper, but extensive numerical evidence suggests
the algorithm is guaranteed to converge under mild conditions
(Section~\ref{sec:sim}).

Unlike the separable covariance model, where the component matrices are only
identified up to a scalar factor, the parameters of the separable correlation
model are identifiable. Standard errors for individual entries of $C_1$, $C_2$,
and $D$ can therefore be obtained from the expected Fisher information. The
computational efficiency of the algorithm also makes parametric bootstrap
likelihood ratio tests of separability practical; simulations show that such
tests maintain nominal size in settings where asymptotic tests fail.

\section{Likelihood-based inference}

\subsection{Maximum likelihood algorithm}\label{sec:mle}
In some applications, including the data example in Section \ref{sec:data-example}, every
element of $Y_i$ is affected by the same vector of predictors $x_i \in
\R{p}$, $i \in \{1, \dots, n\}$. This paper treats this setting for simplicity, but the methods extend to cases where there are different predictors for
different elements.

Assume for now, with $\Sigma$ as in \eqref{eq:sep_cor} and $\Beta \in \R{p\times rc}$ a matrix of regression coefficients, that
\begin{equation}
  \vecop(Y_i) \sim \rN_{rc}(\Beta^\top x_i, \Sigma),\quad i \in \{1, \dots, n\}.
\end{equation}
When $n > p$, because each element has the same vector of predictors, the maximum likelihood estimate of $\Beta$ is the least squares estimate $\hat{\Beta} = (\sum_{i = 1}^n x_i x_i^\top)^{-1} \sum_{i = 1}^n x_i \vecop(Y_i)^\top$. Let $E_i \in \R{r\times c}$ be the matrix of residuals defined by $\vecop(E_i) = \vecop(Y_i) - \hat{\Beta}^\top x_i$. Then the profile log-likelihood $\ell_n(D, C_1, C_2)$ is, up to an additive constant,
\begin{equation} \label{eq:prof_ll}
  -\frac{n}{2}\log \vert D(C_2 \otimes C_1)D\vert - \frac{1}{2}\sum_{i=1}^n\vecop(E_i)^\tsp\{D(C_2 \otimes C_1)D\}^{-1}\vecop(E_i).
\end{equation}
To ensure the function is well-defined in the expressions that follow, let us
identify $D$ with the vector $(d_1, \dots, d_{rc})$ of its diagonal elements and
extend $\ell_n$ to $\R{rc} \times \R{r \times r} \times \R{c \times c}$ by
setting $\ell_n(d_1, \dots, d_{rc}, C_1, C_2) = -\infty$ when at least one of $D$,
$C_1$, and $C_2$ are not positive definite. Then, maximum likelihood corresponds
to the constrained optimization problem
\begin{align} \label{eq:opt_prob}
    \max_{(d_1, \dots, d_{rc}) \in \R{rc}, C_1 \in \R{r \times r}, C_2 \in \R{c \times c}} \ell_n(d_1, \dots, d_{rc}, C_1, C_2)\\
  \text{subject to } d_1, \dots, d_{rc} > 0, C_1\text{ and } C_2\text{ are correlation matrices}. \notag
\end{align}

The proposed approach is a block-coordinate ascent with rescaling, iteratively updating $D$, $C_1$, and $C_2$ until convergence. Letting $(k)$
denote iterations, the strategy is to first find quantities $\check{D}^{(k +
1)}$, $\check{C}_1^{(k + 1)}$, and $\check{C}_2^{(k + 1)}$ which increase the
log-likelihood but which need not be in the feasible set; that is, they need
not satisfy the constraints in \eqref{eq:opt_prob}. Then those quantities are
rescaled to get iterates $D^{(k + 1)}$, $C_1^{(k + 1)}$, and $C_2^{(k + 1)}$
that are in the feasible set and give the same value of the log-likelihood.

At the $k$th iteration, compute, for $j \in \{1, \dots, rc\}$,
\begin{align*}
  \check{d}_j^{(k + 1)} &= \argmax_{d_j > 0} \ell_n(\check{d}_1^{(k + 1)}, \dots, \check{d}_{j - 1}^{(k + 1)}, d_j, d_{j + 1}^{(k)}, \dots d^{(k)}_{rc}, C_1^{(k)}, C_2^{(k)});\\
  \check{C}_1^{(k + 1)} &= \argmax_{C_1 \in \R{r \times r}} \ell_n(\check{D}^{(k + 1)}, C_1, C_2^{(k)});\\
  \check{C}_2^{(k + 1)} &= \argmax_{C_2 \in \R{c \times c}} \ell_n(\check{D}^{(k + 1)}, \check{C}_1^{(k + 1)}, C_2),
\end{align*}
where $\check{D}^{(k + 1)} = \diag(\check{d}_1^{(k + 1)}, \dots,
\check{d}_{rc}^{(k + 1)})$. These updates have closed form expressions, which
leads to a computationally efficient algorithm. For brevity, the solutions are stated here; details are in the Appendix. Define, for $j \in \{1,
\dots, rc\}$ and $k \in \{0, 1, 2, \dots\}$,
\[
  a^{(k+1)}_j = \left\{\left(C_2^{(k)}\otimes C_1^{(k)}\right)^{-1} \tilde{D}_{(-j)}^{(k+1)} S_n\right\}_{jj},
\]
where $S_n = \sum_{i=1}^n \vecop(E_i)\vecop(E_i)^\tsp$ and
$\tilde{D}_{(-j)}^{(k + 1)} = \diag(1/ \check{d}_1^{(k + 1)}, \dots, 1/ \check{d}_{j
- 1}^{(k + 1)}, 0, 1/ d_{j+ 1}^{(k)}, \dots, 1/ d_{rc}^{(k)})$. Also letting $b_j^{(k)} = 4(C_2^{(k)}\otimes C_1^{(k)})^{-1}_{j,j}\{S_n\}_{j,j}$, the maximizer is
\begin{equation} \label{eq:check_d}
  \check{d}_j^{(k + 1)} = \frac{a_j^{(k+1)} + \sqrt{(a_j^{(k+1)})^2 + n b_j^{(k)}}}{2n}.
\end{equation}

Maximizing in $C_1$ and $C_2$ is more straightforward. Define
$\check{E}_i^{(k + 1)}$ by $\vecop(\check{E}_i^{(k + 1)}) =
(\check{D}^{(k + 1)})^{-1}\vecop(E_i)$. Then
\begin{equation} \label{eq:check_C1}
  \check{C}_1^{(k + 1)} =\frac{1}{nc}\sum_{i=1}^n \check{E}_i^{(k + 1)} (C_2^{(k)})^{-1} (\check{E}_i^{(k + 1)})^\tsp.
\end{equation}
and
\begin{equation} \label{eq:check_C2}
  \check{C}_2^{(k + 1)} =\frac{1}{nr}\sum_{i=1}^n (\check{E}_i^{(k + 1)})^\tsp (\check{C}_1^{(k + 1)})^{-1} \check{E}_i^{(k + 1)}.
\end{equation}
The last step of the iteration is the rescaling. Define the matrix of standard deviations $\check{D}^{(k + 1)}_1 = \{I_{r} \circ \check{C}_1^{(k + 1)}\}^{1/2}$, where $\circ$ is the elementwise product; and, similarly, $\check{D}^{(k + 1)}_2 = \{I_{c} \circ \check{C}_2^{(k + 1)}\}^{1/2}$. Then the updates are
\begin{align}
  D^{(k + 1)} &= (\check{D}_2^{(k + 1)} \otimes \check{D}_1^{(k + 1)}) \check{D}^{(k + 1)};\label{eq:d_update} \\
  C_1^{(k + 1)} &= (\check{D}_1^{(k + 1)})^{-1} \check{C}_1^{(k + 1)}(\check{D}_1^{(k + 1)})^{-1}; \label{eq:c1_update} \\
  C_2^{(k + 1)} &= (\check{D}_2^{(k + 1)})^{-1} \check{C}_2^{(k + 1)}(\check{D}_2^{(k + 1)})^{-1}. \label{eq:c2_update}
\end{align}
Routine calculations show these updates are indeed in the feasible set. In
particular, $C_1^{(k + 1)}$ and $C_2^{(k + 1)}$ are correlation matrices by
construction, and $D^{(k + 1)}$ is a diagonal matrix of positive entries.
Moreover, the rescaling does not change the value of the log-likelihood
since the resulting covariance matrix is the same:
\[
  D^{(k + 1)}(C_2^{(k + 1)} \otimes C_1^{(k + 1)})D^{(k + 1)} = \check{D}^{(k + 1)}(\check{C}_2^{(k + 1)} \otimes \check{C}_1^{(k + 1)})\check{D}^{(k + 1)}.
\]
The full algorithm is given in Algorithm \ref{alg:mle}. The following theorem
provides theoretical support for the algorithm. The theorem does not assume
normality; that is, the convergence properties of the algorithm hold even if
fitting the model to non-normal data.
\begin{thm} \label{thm:descent}
  If $n - p \geq \max(r, c)$ and the elements of each $Y_i$ have a joint
  continuous distribution, then the following hold for Algorithm \ref{alg:mle} almost surely:
  (i) $\check{d}_1^{(k + 1)}, \dots, \check{d}_{rc}^{(k + 1)}, \check{C}_1^{(k + 1)}, \check{C}_2^{(k + 1)}$ are unique solutions to their respective
  optimization problem for every $k$; (ii) $d_1^{(k)}, \dots, d_{rc}^{(k)},
  C_1^{(k)}, C_2^{(k)}$ are in the feasible set for every $k$; and (iii) every
  iteration strictly increases the log-likelihood, unless the
  current iterate is a stationary point, in which case the algorithm terminates.
\end{thm}

In practice, the sample size condition is not binding: if the algorithm completes without encountering an indefinite update, then every iterate is in the feasible set and the log-likelihood strictly increased at every iteration, regardless of whether $n - p \geq \max(r, c)$ holds.

For small sample sizes, \eqref{eq:opt_prob} need not have a solution. The
following proposition provides a sample size condition which ensures a solution.
Numerical evidence in Section~\ref{sec:sim} suggests quite a bit smaller sample
sizes are enough for a unique solution to exist, and for Algorithm~\ref{alg:mle}
to converge to it.

\begin{prop} \label{prop:mleexist}
  When $n > p + rc$, the optimization problem \eqref{eq:opt_prob} has a solution with probability one.
\end{prop}

  \begin{algorithm}[ht!]
   \caption{Maximum likelihood with separable correlation} \label{alg:mle}
   \begin{algorithmic}[1]

    \State {\it Input:} $C_1^{(0)} = I_r, C_2^{(0)} = I_c, D^{(0)} = I_{rc}, \ell_n^{(0)} = -\infty, k = 0$, $\mathtt{tol} > 0$

    \While{$\vert \ell_n^{(k)} - \ell_n^{(k - 1)}\vert / \ell_n^{(k - 1)} > \vert \mathtt{tol}\vert$}

    \For{$j = 1, \dots, rc$}
      \State Compute $\check{d}_j^{(k+1)}$ as in \eqref{eq:check_d}
    \EndFor

    \State Compute $\check{C}_1^{(k+1)}$ as in \eqref{eq:check_C1}

    \State Compute $\check{C}_2^{(k+1)}$ as in \eqref{eq:check_C2}

    \State Rescale to get $D^{(k + 1)}$, $C_1^{(k + 1)}$, and $C_2^{(k + 1)}$ as in \eqref{eq:d_update}--\eqref{eq:c2_update}

    \State Compute $\ell_n^{(k+1)} = \ell_n(D^{(k+1)}, C_1^{(k+1)}, C_2^{(k+1)})$

    \State Update $k \leftarrow k + 1$
    \EndWhile
   \end{algorithmic}
  \end{algorithm}

\subsection{Standard errors}\label{sec:se}
The parameters of the separable correlation model are identifiable
(Proposition~\ref{prop:id}), unlike the separable covariance model where
$\Sigma_1$ and $\Sigma_2$ are only identified up to a scalar factor. Standard
errors for individual entries of $C_1$, $C_2$, and $D$ can therefore be obtained
from the expected Fisher information, which can be computed efficiently using
the Kronecker product structure (see the Appendix). The
observed information could also be used, but modern theory for inference on
variances and covariances in linear models suggests that the expected
information is more reliable, especially when the true parameters are close to
the boundary of the parameter space
\citep{ekvall2022confidence,ekvall2026uniform,zhang2025fast}.

\begin{prop} \label{prop:id}
  For any two sets of parameters $(D, C_1, C_2) \neq (D', C_1', C_2')$, it holds
  that $D(C_2 \otimes C_1)D \neq D'(C_2' \otimes C_1')D'$ and, consequently, the
  parameters are identifiable in any family of distributions determined by the
  covariance matrix $\Sigma$.
\end{prop}

\subsection{Testing separable correlation} \label{sec:testing}
Natural hypotheses to test are (a) $H_0$: Separable covariance v.\ $H_A$:
Separable correlation, and (b) $H_0$: Separable correlation v.\ $H_A$:
Unrestricted covariance. For both (a) and (b), parametric
bootstrap likelihood ratio tests are recommended since, as the simulation
results show, classical likelihood ratio tests are in general anti-conservative. Test
(a) requires $n \geq p + rc$ since otherwise the log-likelihood
under the alternative is unbounded, implying the test would reject with
probability one.

The parametric bootstrap proceeds as follows. Using the original dataset,
compute $\hat{\Sigma}_0$ and $\hat{\Beta}$, the maximum likelihood estimates
under the null hypothesis; for test (a) $\hat{\Sigma}_0 = \hat{\Sigma}_{cov}$
and for test (b) $\hat{\Sigma}_0 = \hat{\Sigma}_{cor}$. Generate $B$ independent
datasets, each with $n$ observations, from the model under $H_0$ with $\Sigma =
\hat{\Sigma}_0$ and $\Beta = \hat{\Beta}$. For each simulated dataset $j \in
\{1, \dots, B\}$, compute the maximum likelihood estimates $\hat{\Sigma}^j$ and
$\hat{\Beta}^j$ under the alternative, and the maximum likelihood estimate
$\hat{\Sigma}^j_0$ under the null. For a level $\alpha$ test, reject the null
hypothesis if the likelihood ratio $L(\hat{\Sigma},
\hat{\Beta})/L(\hat{\Sigma}_0, \hat{\Beta})$ obtained from the original data
exceeds the $(1 - \alpha)$th empirical quantile of $\{\xi_1, \dots, \xi_B\}$,
where $\xi_j = L(\hat{\Sigma}^j,
\hat{\Beta}^j)/L(\hat{\Sigma}^j_0, \hat{\Beta}^j)$ and $L(\Beta, \Sigma)$ denotes the multivariate normal
likelihood for $\vecop(Y_1), \dots, \vecop(Y_n)$ with respective means
$\Beta^\tsp x_i$ ($i = 1, \dots, n$) and common covariance matrix $\Sigma$.

\subsection{Regularized estimation}
When $n$ is small relative to $r$, $c$, or $p$, the iterates in
Algorithm~\ref{alg:mle} may not exist or, even if they do, the maximum
likelihood estimator may not exist or have high variance. A natural remedy is a
regularized estimator that maximizes
\begin{align}\label{pen}
  \ell_n(D, C_1, C_2) - \frac{\lambda}{2}\tr(\Sigma^{-1}),\quad \lambda \geq 0.
\end{align}
Because $\Sigma$ is positive definite, $\tr(\Sigma^{-1})$ is the nuclear norm of
$\Sigma^{-1}$; the trace is the sum of eigenvalues, which are the singular
values. This penalty is convenient because it is invariant under the rescaling
step in Algorithm~\ref{alg:mle}; monotone increase of the penalized objective
then follows by the same argument as Theorem~\ref{thm:descent}(iii). Second, the
updates for $D$, $C_1$, and $C_2$ retain closed-form solutions (Appendix). The penalty parameter $\lambda$ can be chosen by
cross-validation or an information criterion.

\section{Data example} \label{sec:data-example}
The method is illustrated using dissolved oxygen concentration data from the
Mississippi River, collected by the US Army Corps of Engineers' Upper
Mississippi River Restoration Program Long Term Resource Monitoring element
\citep{johnson2008status}. The mean structure uses cubic spline regressors in
the year index to remove systematic trends; since the choice of predictors does
not affect estimation of $\Sigma$, it is not discussed further.
The data consist of $n = 21$ years (1994 -- 2002, 2004 -- 2015) of quarterly
measurements from $r = 16$ areas of the Upper Mississippi River. Observations
from winter are excluded since water is typically mostly frozen in the
northernmost sampling areas in winter, so $c = 3$.

Dissolved oxygen measurements are treated as independent among years. In the raw
data there are several measurements for every year, season, and area. For
convenience, the sample means of these measurements are modeled. Let
$Y_{i,j,k}$ ($i = 1, \dots, n$, $j = 1, \dots, 16$, $k = 1, 2, 3$) denote the
sample mean of the dissolved oxygen measurement in season $k$, area $j$, and
year $i$. For the sample means data, the separable correlation model says that
the correlations between mean dissolved oxygen concentrations in spring, summer,
and fall are the same for all areas. Additionally, the correlations between mean
concentrations from different areas are the same in all seasons. Without
imposing additional restrictions, other than independence among years, the
covariance matrix $\Sigma$ would still be of size $48 \times 48$ in this
example, and there are only $n = 21$ years of data. Thus, in this example,
correlation separability is motivated both as a parsimonious parameterization
enabling maximum likelihood estimation, and as a reflection of the
spatio-temporal structure of the data. It is also of scientific interest to
examine whether, in fact, the separable covariance model can be applied to these
data.

The parametric bootstrap test of $H_0$: Separable covariance v.\ $H_A$:
Separable correlation with $B = 10000$ yields a $p$-value less than $0.0001$.
Each bootstrap sample requires fitting both models, so the bootstrap involves
$20000$ fits in total. Using Algorithm~\ref{alg:mle}, this takes approximately
two minutes. With a general-purpose optimizer (BFGS), a single fit takes roughly
$2.8$ seconds, so the same bootstrap would take over $15$ hours. Correlation
separability cannot be tested against an unstructured covariance matrix here
since $n = 21 < 48 = rc$.

The estimate of the season correlation matrix $C_2$ has off-diagonal entries
$-0.02$, $0.10$, and $0.08$, with standard errors $0.055$, $0.054$, and $0.054$
from the expected information (Section~\ref{sec:se}). A Wald test of $H_0\colon
C_2 = I_3$ gives a test-statistic value of $5.96$ ($df = 3$, $p = 0.11$),
indicating only weak evidence against the null hypothesis of no seasonal
correlation.

\begin{figure}[htbp]
\centering
  \includegraphics[width=14cm]{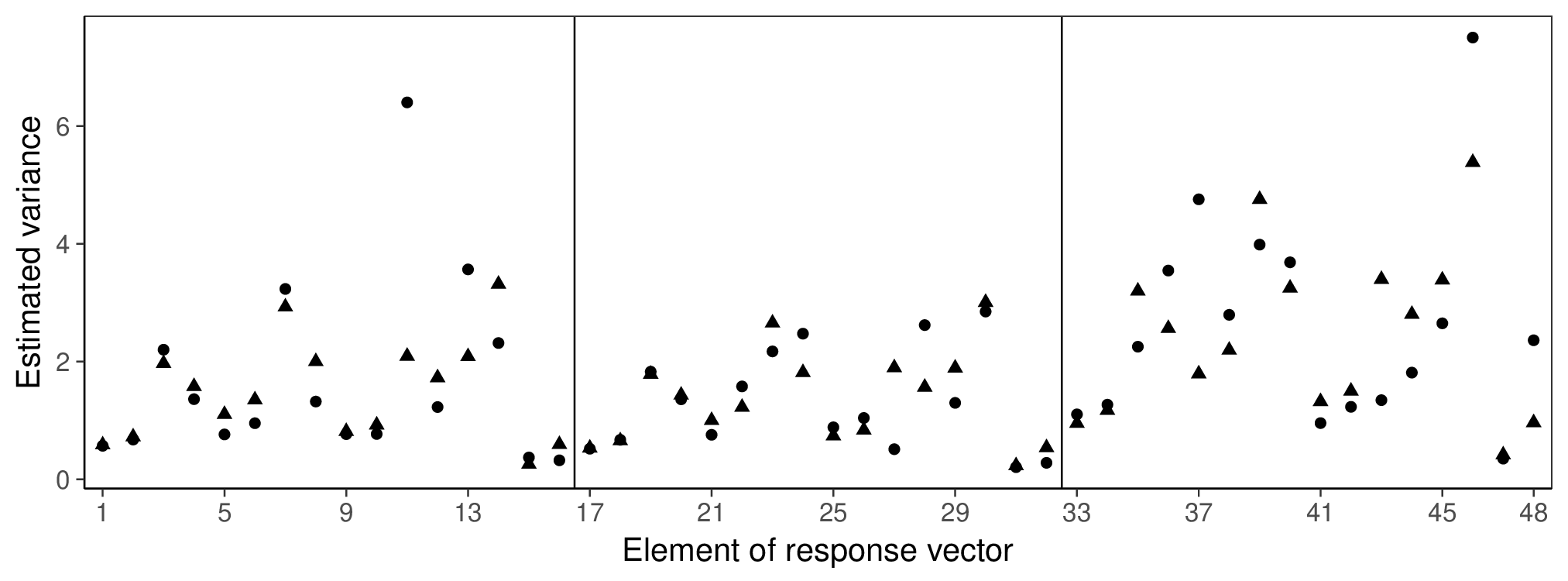}
  \caption{Estimated variances for dissolved oxygen data. Dots and triangles correspond to estimates from separable correlation and separable covariance models, respectively. The three panels are, from left to right, spring, summer, and fall.}
    \label{fig:var}
\end{figure}

The estimated variance of response element 46 is notably large (Fig.\
\ref{fig:var}). This element corresponds to a sampling location with deep water
and low flow, measured in fall when flow is generally lowest (B.\ Gray, personal
communication, Jan 2018). These conditions may cause larger year-to-year
swings in dissolved oxygen, offering a physical explanation. This finding is
inconsistent with separable covariance, which requires any season effect on
variance to be uniform across locations. The variance estimates in Fig.\
\ref{fig:var} reveal other elements where the difference between models is even
greater; for example, separable correlation indicates particularly large
variation at sampling location 11 in spring. Such location-specific findings
would not be detected under separable covariance.
\begin{figure}[htbp]
    \begin{subfigure}[t]{0.5\textwidth}
        \centering
        \includegraphics[width=7cm]{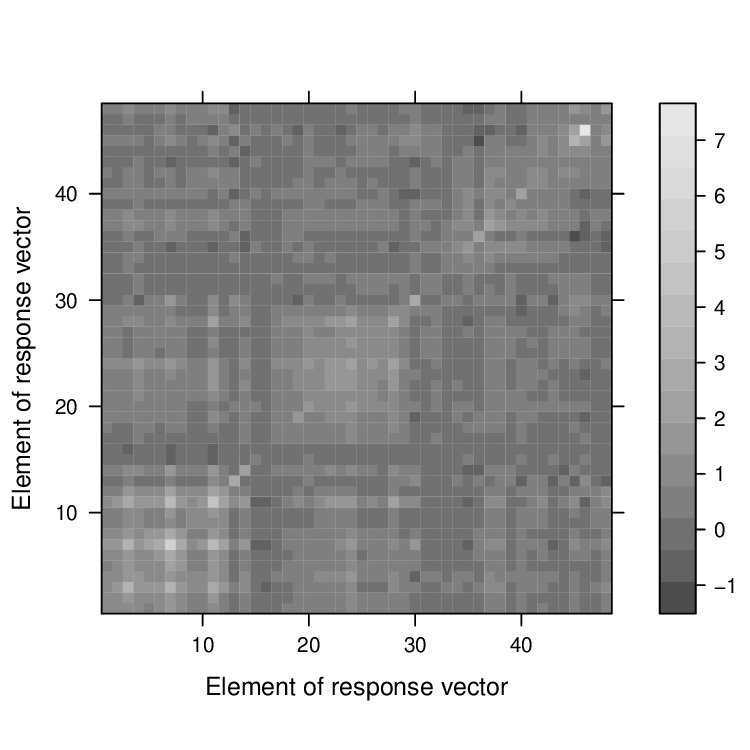}
        \caption{Residual covariance matrix}
    \end{subfigure}%
    ~
    \begin{subfigure}[t]{0.5\textwidth}
        \centering
        \includegraphics[width=7cm]{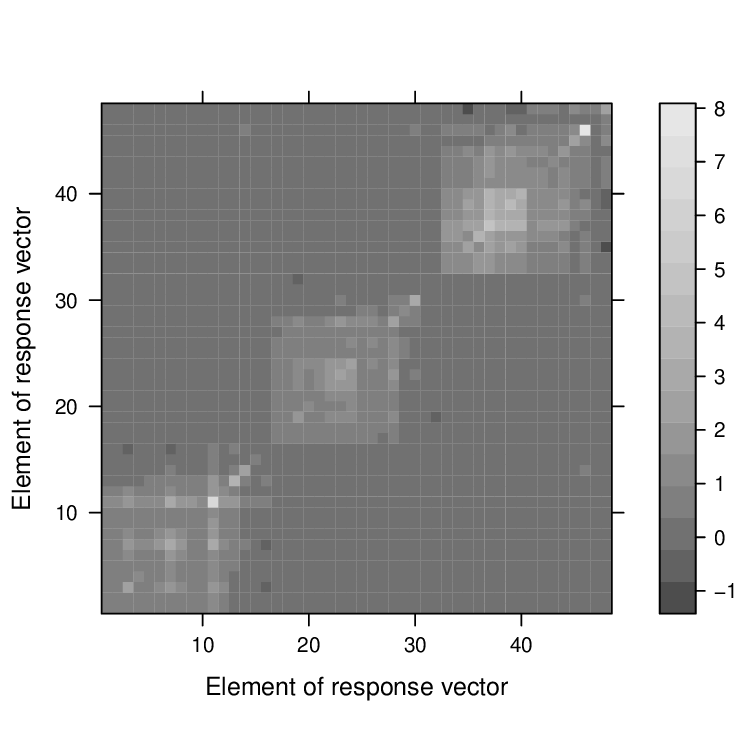}
        \caption{Estimate with separable correlation}
    \end{subfigure}
    \caption{Estimated covariances for dissolved oxygen data. Opacity indicates covariance.}
    \label{fig:cov_heat}
\end{figure}

\section{Simulations} \label{sec:sim}

The parameters for the simulations are taken from the data example,
with $r = 16$ areas, $c = 3$ seasons, and $x_i = 1$, $\beta = 0$. Under the
separable correlation data generating process, $C_1$, $C_2$, and $D$ are the
maximum likelihood estimates from the separable correlation model fit to the
data. Under the separable covariance data generating process, $\Sigma_1$ and
$\Sigma_2$ are the maximum likelihood estimates from the separable covariance
model. The algorithm uses the default convergence tolerance of the
\texttt{sepcor} package ($10^{-8}$) and a maximum of 1000 iterations.

\begin{figure}[htbp]
\centering
\includegraphics[width=\textwidth]{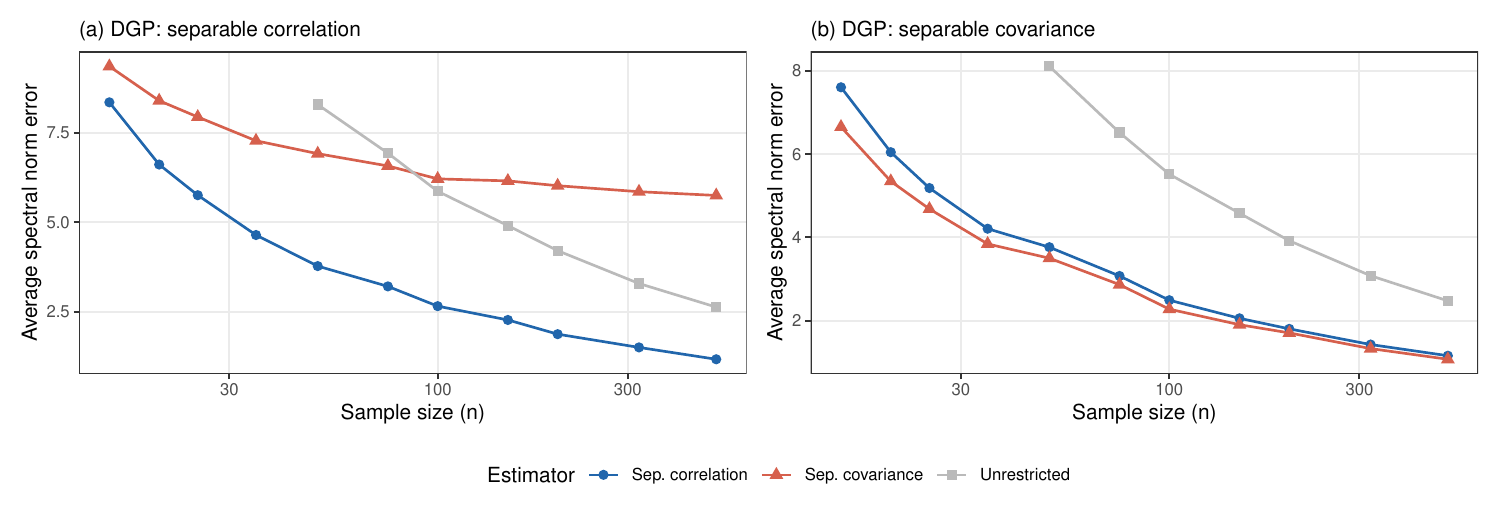}
\caption{Average spectral norm error vs.\ sample size for three estimators, under two data generating processes. Data are generated with parameters set to the estimates from the separable covariance (Panel~(a)) or correlation model (Panel~(b)) in the data example.}
\label{fig:error}
\end{figure}

Figure~\ref{fig:error} shows average spectral norm errors $m^{-1}\sum_{j=1}^m
\|\hat\Sigma^j - \Sigma\|$ across $m = 200$ simulation runs for the separable
correlation, separable covariance, and unrestricted estimators. Under the
separable correlation data generating process, the separable correlation
estimator has lower error than the alternatives at all sample sizes. The
separable covariance estimator has higher error that decreases slowly, due to
persistent bias from incorrectly constraining $D$. Under the separable
covariance data generating process, separable covariance naturally performs
better, but the difference to separable correlation is relatively small. When
covariance is not separable, the unrestricted estimator, available only when $n
> q = 48$, performs better than the separable covariance estimator at large
sample size.

\begin{figure}[htbp]
\centering
\includegraphics[width=\textwidth]{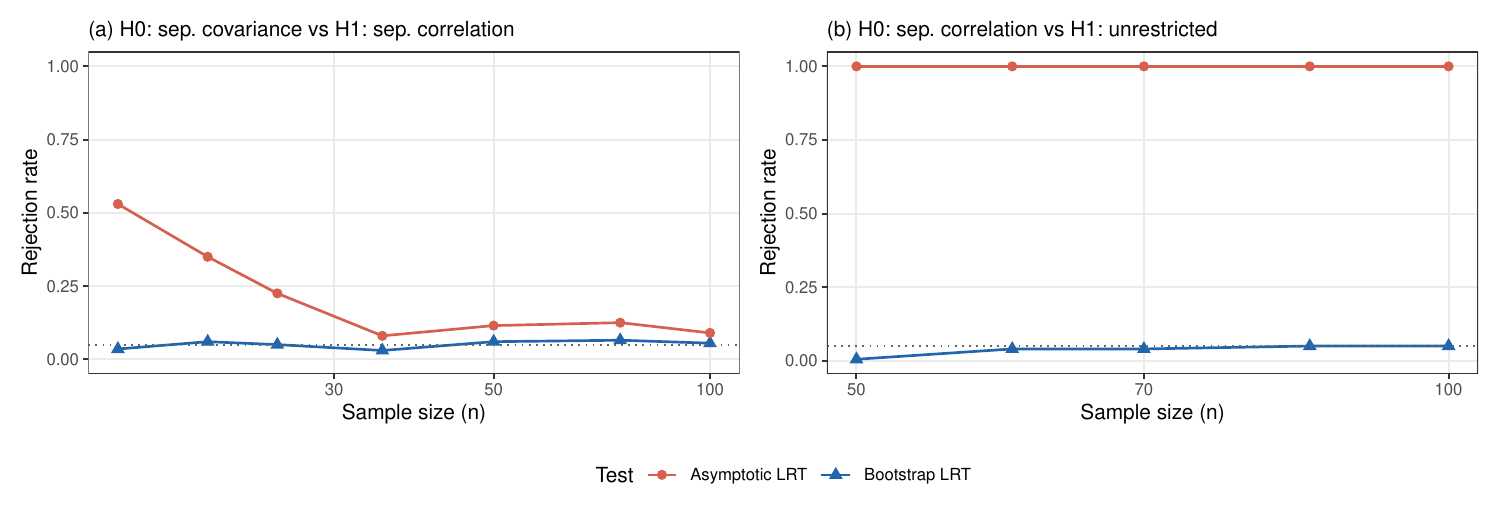}
\caption{Sizes for the bootstrap and asymptotic likelihood ratio tests ($m = 200$ replications, $\alpha = 0.05$). Data are generated with parameters set to the estimates from the separable covariance (Panel~(a)) or correlation model (Panel~(b)) in the data example. The dotted line marks $\alpha = 0.05$.}
\label{fig:tests}
\end{figure}

Figure~\ref{fig:tests} shows rejection rates under the fitted DGPs from the
data example ($r = 16$, $c = 3$). The bootstrap maintains close to nominal
size in both panels. In panel~(a), the asymptotic test over-rejects at small
$n$, exceeding $0.40$ at $n = 15$. In panel~(b), the asymptotic test rejects
in every replication: with $n$ close to $q = 48$, the sample covariance
overfits the data and inflates the test statistic far beyond the
$\chi^2_{1005}$ quantiles. The bootstrap draws from the same small-sample
distribution and controls size.

For the separable covariance model, there has been substantial research on
determining which dimensions ($n, r, c$) ensure, almost surely, (i) a bounded
likelihood, (ii) existence of a maximum likelihood estimator, and (iii)
uniqueness of the maximum likelihood estimator. Here, these questions are
explored empirically for the separable correlation model. Data are generated
from a separable covariance model with autoregressive correlation matrices $[C_1]_{jk} = [C_2]_{jk} = 0.5^{|j - k|}$ and $D = I_{rc}$. For each $(r, c)$, 50 datasets are simulated
and the algorithm is run from 20 random starting values.
Let $\hat{n}_1$ be the smallest $n$ for which all 20 starts converge for all 50
datasets, possibly to different parameter vectors. Let also $\hat{n}_2$ be the
smallest $n$ for which the 20 converged solutions for each dataset agree within
$10^{-4}$ in maximum absolute parameter difference. Then $\hat{n}_1 \leq
\hat{n}_2$, the former being an empirical analogue of almost sure existence and
the latter of almost sure uniqueness.

Table~\ref{tab:boundary} reports these thresholds alongside the exact
theoretical bounds for separable covariance from \citet{derksen2021maximum}. For
the generic case $r \neq c$ with $\gcd(r,c) = 1$, these simplify to $n_1 = n_2 =
\lceil r/c + c/r \rceil$; for $r = c$, $n_1 = 1$ and $n_2 = 3$. The separable
correlation thresholds are consistently larger, reflecting the additional $rc$
variance parameters. For every considered $n \geq \hat{n}_2$, all 20
starting values converge to the same parameter vector, providing
empirical evidence that the maximum likelihood estimator is unique whenever it exists. When $\lambda >
0$, the penalized estimator converges even when the unpenalized algorithm fails;
for example, with $r = 2$, $c = 9$, and $n = 5$, the penalized algorithm with
$\lambda = 1$ always converges.

To assess the coverage of Wald confidence intervals based on the expected Fisher
information, 95\% intervals were computed for each off-diagonal entry of $C_1$ and
$C_2$ across 500 simulation runs with $r = c = 5$, $D = I_{25}$, $[C_1]_{jk} = [C_2]_{jk} = 0.6^{|j - k|}$, and $n \in \{50, 100, 320\}$.
Average empirical coverage rates range from 0.94 to 0.96 across all settings,
within two Monte Carlo standard errors of the nominal level even at $n = 50$.

\begin{table}[h!]
\caption{Sample size thresholds for existence and uniqueness. For separable covariance, $n_1$ and $n_2$ are the exact theoretical bounds from \citet{derksen2021maximum}. For separable correlation, $\hat{n}_1$ and $\hat{n}_2$ are empirical thresholds based on 50 simulated datasets with 20 random starts each.}
\centering
\begin{tabular}{rr cc cc}
 & & \multicolumn{2}{c}{Sep.\ covariance} & \multicolumn{2}{c}{Sep.\ correlation} \\
$r$ & $c$ & $n_1$ & $n_2$ & $\hat{n}_1$ & $\hat{n}_2$ \\
\hline
2 & 2 & 1 & 3 & 4 & 5 \\
2 & 3 & 2 & 2 & 3 & 5 \\
2 & 6 & 3 & 4 & 6 & 6 \\
2 & 9 & 5 & 5 & 8 & 8 \\
3 & 3 & 1 & 3 & 4 & 4 \\
3 & 6 & 2 & 3 & 5 & 6 \\
5 & 5 & 1 & 3 & 5 & 5 \\
5 & 6 & 2 & 2 & 5 & 5 \\
5 & 10 & 2 & 3 & 7 & 7 \\
\end{tabular}
\label{tab:boundary}
\end{table}

Table~\ref{tab:benchmark} compares computing times for the proposed algorithm
against BFGS, where both use the same Kronecker-structured log-likelihood
implemented in \texttt{RcppArmadillo}. Data are generated from a separable correlation model with $[C_1]_{jk} = 0.5^{|j - k|}$, $[C_2]_{jk} = 0.4^{|j - k|}$, and $D = \diag(d_1, \dots, d_{rc})$ with $d_j$ equally spaced from $0.5$ to $2$. The proposed algorithm is 333--2504
times faster across the settings considered, with the advantage growing in $q =
rc$.

\begin{table}[h]
\caption{Mean computing time over 5 replications.}
\centering
\begin{tabular}{rr r rr r}
$r$ & $c$ & $n$ & sepcor (ms) & BFGS (s) & Speedup \\
\hline
5 & 6 &   30 &  0.58 &  0.19 &  333$\times$ \\
8 & 8 &   50 &  1.55 &  0.77 &  499$\times$ \\
6 & 10 &  40 &  1.41 &  0.78 &  555$\times$ \\
10 & 10 & 50 &  2.90 &  2.10 &  723$\times$ \\
8 & 16 &  50 &  4.11 &  4.90 & 1191$\times$ \\
12 & 12 & 50 &  4.99 &  3.01 &  604$\times$ \\
16 & 16 & 50 & 14.23 & 35.63 & 2504$\times$ \\
\end{tabular}
\label{tab:benchmark}
\end{table}
All simulations and timing comparisons were run sequentially (single-threaded) on an Apple M1 Max (3.2\,GHz) with 32\,GB of RAM, using R 4.5.2.

\section{Final remarks}

The proposed algorithm extends naturally to tensor-valued responses, where the
correlation matrix factors as a Kronecker product of $K > 2$ matrices, $C_K
\otimes \cdots \otimes C_1$. The coordinate ascent structure carries over:
cycle through each $C_k$, update over the larger space of positive definite
matrices, and rescale to a correlation matrix. The nuclear norm penalty
\eqref{pen} generalizes similarly.

If the elements of each $Y_i$ need not have the same predictors, say
$\E(Y_{ijk}) = x_{ijk}^\tsp \beta$ for $x_{ijk}, \beta \in \R{p}$, then the
maximum likelihood estimator for $\beta$ in general depends on the maximum
likelihood estimator for $\Sigma$. The proposed algorithm can then be adapted by
including an additional step to update the regression coefficients $\beta$ after
updating $D$, $C_1$, and $C_2$. The update for $\beta$ has a familiar,
generalized least squares-type closed-form solution, and the monotone increase
property of the algorithm is preserved by the same argument as in
Theorem~\ref{thm:descent}(iii).

It is straightforward to impose
restrictions such as $C_1 = I_r$ by skipping the update of $C_1$, effectively fixing
$C_1^{(k + 1)} = I_r$ at every iteration. This would be a useful model for
spatial data with no temporal correlation, for example. Such a restriction can
be tested in the same way as described in Section~\ref{sec:testing}.

In the data example, further years of sampling would permit formal testing of
the validity of the assumed separability against an unstructured covariance
matrix. Restrictions on either the spatial or temporal
autocorrelation could also be incorporated in such a study, building on the work
of \citet{szczepanska-alvarez2017estimation}.

The computational advantages of the proposed algorithm are in line with the best
algorithms for the separable covariance model: \citet{lu2005likelihood} report
that the flip-flop algorithm for the separable covariance model, which is also a
block-coordinate descent algorithm \citep{dutilleul1999mle} is 50 to 5000 times
faster than Newton--Raphson for separable covariance, and the gap grows with
dimension.

Theoretical results on the existence and uniqueness of the maximum likelihood estimator for the
separable correlation model are an avenue for future research. For example, it
would be of interest to explore whether the likelihood is pathwise concave along
some appropriate path, as it is along geodesics in the space of positive
definite matrices for the separable covariance model.

Further work on regularized estimation may also be of interest. In addition to
exploring other penalties than the one considered here, it may be possible to
extend the core shrinkage of \citet{hoff2023core} to the separable correlation
model.

\section*{Acknowledgments}
The author thanks Brian Gray, Aaron Molstad, and Daniel Eck for discussions and suggestions that
led to the improvement of this article. Ekvall gratefully acknowledges support
by the American--Scandinavian Foundation.

\section*{Supplementary material}
The supplementary material includes code for reproducing the data example and simulations, and is available at \url{https://github.com/koekvall/sepcor-suppl}.

\bibliographystyle{apalike}
\bibliography{sepcor.bib}

\begin{appendices}

\section{Proofs} \label{app:proofs}
\begin{proof}[Proof of Proposition \ref{prop:id}]
  Let $\Sigma = D(C_2 \otimes C_1)D$ and $\Sigma' = D'(C_2' \otimes C_1')D'$. If
  $D \neq D'$, then $\Sigma$ and $\Sigma'$ have different diagonals, since the
  diagonal entries of $\Sigma$ are the squares of those of $D$.

  If $D = D'$ but $C_1 \neq C_1'$, then $\Sigma = \Sigma'$ requires
  $C_2 \otimes C_1 = C_2' \otimes C_1'$ since $D$ is invertible. The upper-left
  $r \times r$ block of $C_2 \otimes C_1$ is $[C_2]_{11} C_1 = C_1$, and
  similarly for $C_2' \otimes C_1'$, so $C_1 \neq C_1'$ gives a contradiction.

  If $D = D'$ and $C_1 = C_1'$ but $C_2 \neq C_2'$, then $C_2 \otimes C_1 = C_2'
  \otimes C_1$ implies $C_2 \otimes I_r = C_2' \otimes I_r$ upon right- and
  left-multiplying by $I_c \otimes C_1^{-1}$, contradicting $C_2 \neq C_2'$.
\end{proof}

\begin{lemma}\label{lem:rank}
  Let $E_i \in \R{r \times c}$ be the residual matrices defined in Section~\ref{sec:mle}, with $n - p \geq \max(r/c, c/r)$, and assume the joint distribution of $Y_1, \dots, Y_n$ is absolutely continuous. Then, for any positive definite $B$, almost surely,
  \[
    \operatorname{rank}\!\left(\sum_{i=1}^n E_i B E_i^\tsp\right) = r;
  \]
  for any positive definite $A$, almost surely,
  \[
    \operatorname{rank}\!\left(\sum_{i=1}^n E_i^\tsp A E_i\right) = c;
  \]
  and all diagonal elements of $S_n = \sum_{i=1}^n \vecop(E_i)\vecop(E_i)^\tsp$ are almost surely positive.
\end{lemma}

\begin{proof}
  Let $\mathcal{Y} = [\vecop(Y_1), \dots, \vecop(Y_n)]^\tsp \in \R{n\times rc}$ and $X = [x_1, \dots, x_n]^\tsp \in \R{n \times p}$.

  Let $M = I_n - X(X^\tsp X)^{-1}X^\tsp$. Define $E \in \R{nc \times r}$
  by stacking $E_1^\tsp, \dots, E_n^\tsp$ vertically, so $E = [E_1, \dots, E_n]^\tsp$; and similarly
  $\tilde{Y} \in \R{nc \times r}$ by stacking
  $Y_1^\tsp, \dots, Y_n^\tsp$. Then
  $E = (M \otimes I_c)\tilde{Y}$ and
  \[
    \sum_{i=1}^n E_i B E_i^\tsp = E^\tsp (I_n \otimes B) E.
  \]
  Since $I_n \otimes B$ is positive definite, the null spaces of
  $E^\tsp(I_n \otimes B)E$ and $E^\tsp E$ coincide. In particular,
  $\operatorname{rank}(\sum_{i=1}^n E_i B E_i^\tsp) = \operatorname{rank}(E)$
  for every positive definite $B$. It suffices to show
  $\operatorname{rank}(E) = r$.

  Write $M = UU^\tsp$ with $U \in \R{n \times (n-p)}$ having orthonormal
  columns. Then $E = (U \otimes I_c)H$, where
  $H = (U^\tsp \otimes I_c)\tilde{Y} \in \R{(n-p)c \times r}$.
  Since $U \otimes I_c$ has full column rank,
  $\operatorname{rank}(E) = \operatorname{rank}(H)$.

  Now $\vecop(H) = (I_r \otimes U^\tsp \otimes I_c)\vecop(\tilde{Y})$.
  The matrix $I_r \otimes U^\tsp \otimes I_c$ has full row rank $(n-p)cr$, so
  this linear map from $\R{nrc}$ to $\R{(n-p)cr}$ is onto. Since the entries
  of $\tilde{Y}$ are those of $Y_1, \dots, Y_n$, the distribution
  of $\vecop(\tilde{Y})$ is absolutely continuous, and therefore so is that
  of $H$. The set of $(n-p)c \times r$ matrices with rank below
  $\min((n-p)c, r)$ is a proper algebraic subset, hence has Lebesgue measure
  zero. Therefore $\operatorname{rank}(H) = \min((n-p)c, r)$ almost surely,
  and since $n - p \geq r/c$ gives $(n-p)c \geq r$, it follows that
  $\operatorname{rank}(H) = r$ almost surely. The same argument with $r$
  and $c$ interchanged shows
  $\operatorname{rank}\!\left(\sum_{i=1}^n E_i^\tsp A E_i\right) = c$
  almost surely, using $(n-p)r \geq c$ from $n - p \geq c/r$.

  For the diagonal entries of $S_n$, let $\mathcal{Y} \in \R{n \times rc}$ be
  the data matrix with $i$-th row $\vecop(Y_i)^\tsp$ and
  $\mathcal{E} = M\mathcal{Y}$. Then $[S_n]_{jj} = \sum_{i=1}^n
  [\vecop(E_i)]_j^2$ is the squared norm of the $j$-th column of
  $\mathcal{E}$, which is $My_{\cdot j}$, where $y_{\cdot j} \in \R{n}$
  collects the $j$-th entry of $\vecop(Y_i)$ across $i = 1, \dots, n$.
  Since $M$ has rank $n - p \geq 1$, its null space $\operatorname{col}(X)$ is
  a proper subspace of $\R{n}$, so the event $My_{\cdot j} = 0$ has Lebesgue
  measure zero. The marginal distribution of $y_{\cdot j}$ is absolutely
  continuous, so $[S_n]_{jj} > 0$ almost surely.
\end{proof}

\begin{proof}[Proof of Theorem \ref{thm:descent}]
The proof proceeds in order. Let $Q_X = I_n - X(X^\tsp X)^{-1}X^\tsp$ and $\mathcal{Y} \in \R{n \times rc}$ have $i$th row $\vecop(Y_i)^\tsp$, so that $\mathcal{E} = Q_X\mathcal{Y}$ has $i$th row $\vecop(E_i)^\tsp$.

\textit{Part (i).} The argument proceeds by induction on $k$. The base case $k = 0$ holds because $C_1^{(0)} = I_r$ and $C_2^{(0)} = I_c$ are positive definite and $D^{(0)} = I_{rc}$ is in the feasible set. For the inductive step, assume $C_1^{(k)}$ and $C_2^{(k)}$ are positive definite and let $A^{(k)} = (C_2^{(k)}\otimes C_1^{(k)})^{-1}$. With all parameters except $d_j$ fixed, the $d_j$-dependent terms of $\ell_n$ are $h(d_j) = -n\log d_j - b_j^{(k)}/(8d_j^2) - a_j^{(k+1)}/d_j$, where the middle term uses $b_j^{(k)} = 4A^{(k)}_{jj}[S_n]_{jj}$ and the last term collects the cross terms $A^{(k)}_{jm}[S_n]_{jm}/d_m$ for $m \neq j$ (using symmetry of $A^{(k)}$ and $S_n$). The quantity $[S_n]_{jj} > 0$ almost surely by Lemma~\ref{lem:rank}. Since $A^{(k)}_{jj} > 0$, both $b_j^{(k)} > 0$ and $h(d_j) \to -\infty$ as $d_j \to 0^+$ or $d_j \to \infty$, so $h$ attains its maximum on $(0,\infty)$. Setting $h'(d_j) = 0$ and multiplying by $-d_j^3$ gives
\[
  nd_j^2 - a_j^{(k+1)} d_j - \tfrac{1}{4}b_j^{(k)} = 0,
\]
a quadratic with positive discriminant $(a_j^{(k+1)})^2 + nb_j^{(k)} > 0$ and unique positive root $(a_j^{(k+1)} + \sqrt{(a_j^{(k+1)})^2 + nb_j^{(k)}})/(2n)$, confirming \eqref{eq:check_d}.

For $\check{C}_1^{(k+1)}$, the objective over positive definite matrices $C_1$ is, up to constants,
\[
  g_1(C_1) = -\frac{nc}{2}\log\det(C_1) - \frac{1}{2}\operatorname{tr}\!\left(C_1^{-1}T_1^{(k)}\right),
\]
where $T_1^{(k)} = \sum_{i=1}^n \check{E}_i^{(k+1)} (C_2^{(k)})^{-1}(\check{E}_i^{(k+1)})^\tsp$. Routine calculations show that $g_1$ is strictly concave in $C_1^{-1}$, and hence has unique maximizer $T_1^{(k)}/(nc)$ provided $T_1^{(k)}$ is positive definite. This holds almost surely, as shown next.

Define $\check{E} = [\check{E}_1^{(k+1)}, \dots, \check{E}_n^{(k+1)}]^\tsp \in \R{nc \times r}$ by stacking $(\check{E}_i^{(k+1)})^\tsp$ vertically. Then $T_1^{(k)} = \check{E}^\tsp(I_n \otimes (C_2^{(k)})^{-1})\check{E}$. Since $(C_2^{(k)})^{-1}$ is positive definite, the null spaces of $T_1^{(k)}$ and $\check{E}^\tsp \check{E}$ coincide, so $\operatorname{rank}(T_1^{(k)}) = \operatorname{rank}(\check{E})$. It remains to show $\operatorname{rank}(\check{E}) = r$ almost surely.

Partition $\check{E}$ into $c$ blocks: the $b$th block $\check{E}^{(b)} \in \R{n \times r}$ has $(i, a)$ entry $[E_i]_{a,b}/\check{d}_{(b-1)r+a}^{(k+1)}$, so $\check{E}^{(b)} = E^{(b)} D_b^{-1}$, where $E^{(b)} \in \R{n \times r}$ has $(i,a)$ entry $[E_i]_{a,b}$ and $D_b = \diag(\check{d}_{(b-1)r+1}^{(k+1)}, \dots, \check{d}_{br}^{(k+1)})$. Since each $\check{d}_j^{(k+1)} > 0$ almost surely, $D_b$ is invertible and $\operatorname{rank}(\check{E}^{(b)}) = \operatorname{rank}(E^{(b)})$. The columns of $E^{(b)}$ are $r$ columns of $\mathcal{E} = Q_X\mathcal{Y}$, so by the argument in Lemma~\ref{lem:rank}, $\operatorname{rank}(E^{(b)}) = \min(n - p, r) = r$ almost surely; the last equality uses $n - p \geq r$. Since $\check{E}^{(b)}$ is a row-submatrix of $\check{E}$ with $r$ linearly independent columns, $\operatorname{rank}(\check{E}) = r$ almost surely. Note that Lemma~\ref{lem:rank} only requires $n - p \geq r/c$ because it uses all $nc$ rows of the stacked matrix, whereas the single-block argument here uses only $n$ rows and therefore requires $n - p \geq r$.

For $\check{C}_2^{(k+1)}$, the objective has the same form with maximizer $T_2^{(k)}/(nr)$, where $T_2^{(k)} = \sum_{i=1}^n (\check{E}_i^{(k+1)})^\tsp (\check{C}_1^{(k+1)})^{-1} \check{E}_i^{(k+1)}$. Since $\check{C}_1^{(k+1)}$ is positive definite almost surely, the null space argument reduces the rank of $T_2^{(k)}$ to that of the $nr \times c$ matrix $F$ obtained by stacking $\check{E}_1^{(k+1)}, \dots, \check{E}_n^{(k+1)}$ vertically. Partition $F$ into $r$ blocks: the $a$th block $F^{(a)} \in \R{n \times c}$ satisfies $F^{(a)} = G^{(a)}\tilde{D}_a^{-1}$, where $G^{(a)} \in \R{n \times c}$ has $(i,b)$ entry $[E_i]_{a,b}$ and $\tilde{D}_a = \diag(\check{d}_a^{(k+1)}, \check{d}_{r+a}^{(k+1)}, \dots, \check{d}_{(c-1)r+a}^{(k+1)})$. By the argument in Lemma~\ref{lem:rank}, $\operatorname{rank}(G^{(a)}) = \min(n - p, c) = c$ almost surely, using $n - p \geq c$. Since $\tilde{D}_a$ is invertible and $F^{(a)}$ is a row-submatrix of $F$, $\operatorname{rank}(F) = c$ and $T_2^{(k)}$ is positive definite almost surely.

\textit{Part (ii).} Consider $C_1^{(k+1)}$ from \eqref{eq:c1_update}. Since $\check{C}_1^{(k+1)}$ is positive definite (part (i)) and $\check{D}_1^{(k+1)} = (I_r\circ \check{C}_1^{(k+1)})^{1/2}$ is invertible, $C_1^{(k+1)} = (\check{D}_1^{(k+1)})^{-1}\check{C}_1^{(k+1)}(\check{D}_1^{(k+1)})^{-1}$ is positive definite. Its diagonal entries are $[C_1^{(k+1)}]_{jj} = [\check{C}_1^{(k+1)}]_{jj}/[\check{C}_1^{(k+1)}]_{jj} = 1$, so $C_1^{(k+1)}$ is a correlation matrix. The argument for $C_2^{(k+1)}$ is identical. That $D^{(k+1)}$ is a diagonal matrix with positive entries follows directly from \eqref{eq:d_update} and the positivity of $\check{D}^{(k+1)}$, $\check{D}_1^{(k+1)}$, and $\check{D}_2^{(k+1)}$.

\textit{Part (iii).} Each coordinate update weakly increases $\ell_n$ by definition of the argmax, and strictly so unless the current iterate already maximizes that coordinate. The rescaling step \eqref{eq:d_update}--\eqref{eq:c2_update} leaves $D(C_2\otimes C_1)D$ unchanged, hence $\ell_n$ unchanged. The algorithm terminates when no coordinate update strictly increases $\ell_n$, i.e., at a stationary point.
\end{proof}

\section{Penalized updates}\label{app:pen}
Let $\bar{D} \in \R{r\times c}$ be the matrix with $(j_1, j_2)$ entry equal to the
standard deviation of response element $(j_1, j_2)$, i.e., $[\bar{D}]_{j_1 j_2} =
d_{j_1 + (j_2-1)r}$. For fixed $\check{D}^{(k+1)}$ and $C_2^{(k)}$, the penalty
satisfies $\tr(\Sigma^{-1}) = \tr(W_r^{(k)} C_1^{-1})$, where
$W_r^{(k)} = \diag(\check{\bar{D}}^{-2}\,\diag(C_2^{(k)})^{-1})$ is $r\times r$ diagonal
and $\check{\bar{D}}^{-2}$ denotes elementwise squaring and inverting of $\check{\bar{D}}^{(k+1)}$.
For fixed $\check{D}^{(k+1)}$ and $\check{C}_1^{(k+1)}$,
$\tr(\Sigma^{-1}) = \tr(W_c^{(k)} C_2^{-1})$, where
$W_c^{(k)} = \diag((\check{\bar{D}}^{-2})^\tsp\,\diag(\check{C}_1^{(k+1)})^{-1})$ is $c\times c$ diagonal.
The penalized updates replace \eqref{eq:check_C1}--\eqref{eq:check_C2} with
\begin{align}
  \check{C}_1^{(k+1)} &= \frac{1}{nc}\!\left[\sum_{i=1}^n \check{E}_i^{(k+1)} (C_2^{(k)})^{-1} (\check{E}_i^{(k+1)})^\tsp + \lambda W_r^{(k)}\right], \label{eq:check_C1_pen}\\
  \check{C}_2^{(k+1)} &= \frac{1}{nr}\!\left[\sum_{i=1}^n (\check{E}_i^{(k+1)})^\tsp (\check{C}_1^{(k+1)})^{-1} \check{E}_i^{(k+1)} + \lambda W_c^{(k)}\right], \label{eq:check_C2_pen}
\end{align}
and the update for $\check{d}_j$ in \eqref{eq:check_d} is unchanged in form, with
$[S_n]_{jj}$ replaced by $[S_n]_{jj} + \lambda/n$. The rescaling step
\eqref{eq:d_update}--\eqref{eq:c2_update} is unchanged.

\section{Fisher information}\label{app:fisher}
Without loss of generality, assume that $\Beta = 0$ since the score for the covariance parameters depends on $\Beta$ only through
$\vecop(Y_i) - \Beta^\top x_i \sim \rN(0, \Sigma)$, $i \in \{1,\dots,n\}$. In more general
notation, let $\Sigma$ be parameterized by $\theta = (\theta_1, \dots,
\theta_m)$, where $\theta_j$ is either an off-diagonal entry of $C_1$ or $C_2$,
or a diagonal entry of $D$. The log-likelihood is, up to a constant,
\[
  \ell_n(\theta) = -\frac{n}{2}\log\lvert\Sigma\rvert - \frac{1}{2}\sum_{i=1}^n \vecop(Y_i)^\tsp \Sigma^{-1} \vecop(Y_i).
\]
With $H_j = \partial\Sigma/\partial\theta_j$, the score is
\[
  \frac{\partial\ell_n}{\partial\theta_j} = -\frac{n}{2}\tr(\Sigma^{-1}H_j) + \frac{1}{2}\sum_{i=1}^n \vecop(Y_i)^\tsp \Sigma^{-1}H_j\Sigma^{-1} \vecop(Y_i).
\]
The Fisher information has elements $\mathcal{I}_{jk} =
\cov(\partial\ell_n/\partial\theta_j,\, \partial\ell_n/\partial\theta_k)$. The
first term in the score is constant in the data and does not contribute;
independence of observations gives
\[
  \mathcal{I}_{jk} = \frac{n}{4}\cov\!\left(\vecop(Y_1)^\tsp A_j \vecop(Y_1),\; \vecop(Y_1)^\tsp A_k \vecop(Y_1)\right),
\]
where $A_j = \Sigma^{-1}H_j\Sigma^{-1}$ and $\vecop(Y_1) \sim \rN(0, \Sigma)$. For a random vector $v \sim \rN(0,\Sigma)$ and
symmetric matrices $A$, $B$, the identity $\cov(v^\tsp Av,\, v^\tsp Bv) = 2\tr(A\Sigma B\Sigma)$
yields
\[
  \cov\!\left(\vecop(Y_1)^\tsp A_j \vecop(Y_1),\; \vecop(Y_1)^\tsp A_k \vecop(Y_1)\right) = 2\tr(\Sigma^{-1}H_j\Sigma^{-1}H_k),
\]
and therefore $\mathcal{I}_{jk} = \tfrac{n}{2}\tr(\Sigma^{-1}H_j\Sigma^{-1}H_k)$.

Computation of the information matrix is efficient because $\Sigma^{-1} = D^{-1}(C_2^{-1}\otimes C_1^{-1})D^{-1}$
and each $H_j$ is sparse or has Kronecker structure. For $\theta_j = [C_1]_{ab}$, the derivative
is $H_j = D\{C_2\otimes(E_{ab}+E_{ba})\}D$, where $E_{ab}$ is the elementary matrix with a one
in position $(a,b)$ and zeros elsewhere. The mixed-product property gives
$\Sigma^{-1}H_j = D^{-1}\{I_c\otimes C_1^{-1}(E_{ab}+E_{ba})\}D$,
so for $\theta_k = [C_1]_{cd}$,
\[
  \tr(\Sigma^{-1}H_j\Sigma^{-1}H_k) = c\cdot\tr\!\left\{C_1^{-1}(E_{ab}+E_{ba})C_1^{-1}(E_{cd}+E_{dc})\right\}.
\]
For a cross pair $\theta_j = [C_1]_{ab}$ and $\theta_k = [C_2]_{cd}$, the trace
factors as $\tr\{C_1^{-1}(E_{ab}+E_{ba})\}\cdot\tr\{C_2^{-1}(E_{cd}+E_{dc})\}$. For $\theta_j = d_k$, $H_j = e_ke_k^\top(C_2\otimes C_1)D +
D(C_2\otimes C_1)e_ke_k^\top$ has rank two, so the trace reduces to a sum of
$O(rc)$ entries of $\Sigma^{-1}$. In each case, the $rc\times rc$ matrices need
not be formed explicitly.

\end{appendices}

\end{document}